\documentclass[conference]{IEEEtran}
\usepackage{amsmath}
\usepackage{color}
\usepackage{array}
\usepackage{stfloats}
\usepackage{amsthm} 
\usepackage{amssymb}
\usepackage{float}
\usepackage{nccmath}
\usepackage{graphicx}
\usepackage[linesnumbered,ruled]{algorithm2e}
\usepackage{setspace}
\usepackage{booktabs}
\usepackage{subcaption}
\usepackage{mwe}
\usepackage{cite}
\usepackage{amsmath,amssymb,amsfonts}
\usepackage{graphicx}
\usepackage{textcomp}
\usepackage{xcolor}
\usepackage[utf8]{inputenc} % use UTF8 encoding
\usepackage{kotex} % use KoTeX package for Korean 

\usepackage{algorithmicx}
\usepackage{xspace}
\begin{document}
\newcommand{\BfPara}[1]{{\noindent\bf#1.}\xspace}
\newcommand\mycaption[2]{\caption{#1\newline\small#2}}
\newcommand\mycap[3]{\caption{#1\newline\small#2\newline\small#3}}

\title{Multi-Agent Deep Reinforcement Learning for Cooperative Connected Vehicles}

	\author{\IEEEauthorblockN{Dohyun Kwon$^{\ddag}$ and Joongheon Kim$^{\S}$}
		\IEEEauthorblockA{
			$^{\ddag}$School of Computer Science and Engineering, Chung-Ang University, Seoul, Republic of Korea\\
			$^{\S}$School of Electrical Engineering, Korea University, Seoul, Republic of Korea\\
			E-mails: 
			$^{\ddag}$\texttt{kdh1102@cau.ac.kr}, 
			$^{\S}$\texttt{joongheon@korea.ac.kr}
		}
	}

\maketitle

\begin{abstract}
Millimeter-wave (mmWave) base station can offer abundant high capacity channel resources toward connected vehicles so that quality-of-service (QoS) of them in terms of downlink throughput can be highly improved. The mmWave base station can operate among existing base stations (e.g., macro-cell base station) on non-overlapped channels among them and the vehicles can make decision what base station to associate, and what channel to utilize on heterogeneous networks. Furthermore, because of the non-omni property of mmWave communication, the vehicles decide how to align the beam direction toward mmWave base station to associate with it. However, such joint problem requires high computational cost, which is NP-hard and has combinatorial features. In this paper, we solve the problem in 3-tier heterogeneous vehicular network (HetVNet) with multi-agent deep reinforcement learning (DRL) in a way that maximizes expected total reward (i.e., downlink throughput) of vehicles. The multi-agent deep deterministic policy gradient (MADDPG) approach is introduced to achieve optimal policy in continuous action domain.
% such joing problem ~ requires 사이 생략: of base station association, channel selection, and occasional beam alignment toward mmWave base station 

\end{abstract}

% Introduction
% contents := Connected vehicle (V2I communication) 
% --> 제목을 connected vehicles라고하면 V2V 통신도 포함될 수 있는 것으로 오해할 수 있으므로 V2I로 명시하는 것이 더 나을 수 있음. --> 건의.
%100m 500m 3000m
\section{Introduction}\label{sec:sec1}
% Intro - 1 : HetVNet (multi-tier wireless networks)에 대한 트렌드 및 예상되는 문제점들 언급
The vehicle-to-infrastructure (V2I) communication via millimeter-wave (mmWave) is vital for the successful operation of next generation fifth-generation (5G) intelligent transportation system (ITS)~\cite{Access_Survey}. Moreover, as the demand of wireless spectrum in ITS is enormously increased nowadays, roadside mmWave base stations that configure small cell coverage regions are deployed in multi-tier heterogeneous vehicular networks (HetVNets) to enhance the spectrum efficiency as well as offload the huge traffic burden~\cite{WC_Campolo}. The additional dense deployment of the small cells on HetVNet meets the requirements of ITS vehicles with high data rate and enlarged coverage region. However, as the number of base stations and vehicular user equipment systems (VUEs) is getting dramatically increased, the radio access technology (RAT) of VUEs on HetVNet is challenging to optimize the wireless resource utilization~\cite{WC_Campolo}. The radio resource management in HetVNet for improving the QoS of VUEs is NP-hard and computationally intractable~\cite{CST_Zheng}. Furthermore, considering the fact that the propagation property of mmWave wireless channels, which is highly directive and is only available within short range (i.e., approximately a hundred meter), plenteous mmWave base stations should be densely deployed to support mmWave wireless communication services~\cite{ref1,ref2,ref3,ref4,ref5}. Therefore, decision making with respect to association, channel selection, and occasional beam alignment task of VUEs imposes a heavy computational burden to ensure QoS-aware wireless communication on HetVNet~\cite{TCOM_Rappaport}.

% Intro - 2 : HetVNet 에서의 optimization에 대한 기존 연구들 언급
There have been many research results regarding cell association and resource allocation problem that is called CARA. The resource allocation problem of mmWave-enabled network was considered in~\cite{Globecom_Busari} and~\cite{PIMRC_Shi}. In addition, joint CARA problem was studied in~\cite{WCL_Liu, TWC_Wang, arXiv_Kuang, JSAC_Lin}. However, because of the NP-hard and combinatorial features of joint CARA problem, it is challenging to achieve a globally optimal solution. There have been some approaches to solve the CARA problem, such as graph theory approach~\cite{TVT_Chen}, integer programming method~\cite{PMC_Ortin}, matching game solution~\cite{TVT_LeAnh}, and stochastic geometric strategy~\cite{Infocom_Bao}. These approaches still were limited to solve the joint CARA problem as they needed nearly precise information such as full knowledge of channel state information (CSI) or fading models. In practice, such accurate information may not be available so that computing the optimal point of joint CARA problem is intractable. For this regard, the multi-agent DRL approach is proposed to solve the joint CARA problem in a way that improves the downlink (DL) throughput of VUEs in HetVNet.

% Intro - 3 : 이러한 상황을 종합했을 때, 강화학습이 multi-tier HetVNet에서 CARA문제를 어떻게 효과적으로 해결할 수 있는지 기존 강화학습 기반 solution들을 인용하면서 제시할 것.
The reinforcement learning has been widely applied to solve various types of complex decision making problems in wireless networks such as interference alignment in cache-enabled networks~\cite{TVT_He} and dynamic duty cycle selection technique in unlicensed spectrum~\cite{WCNC_Rupasinghe}. Unlike the existing approaches, the reinforcement learning needs only a few information to operate, such as the possible action space of learning agent. Based on the interaction between agent and its own environment, the reinforcement learning agent observes state transition and learns how to act good by updating its \textit{policy}~\cite{arXiv_Arulkumaran}. The agent estimates the expected total reward per possible actions for given state and make a decision how to act on a sequential decision making process. In~\cite{ICC_Xu}, a power-efficient resource allocation framework for cloud radio access networks (RANs) is proposed based on DRL. Elsayed \textit{et al.}~\cite{GlobeCom_Elsayed} proposed a DRL based latency reducing scheme of mission critical services for the next generation wireless networks. They combined the long short-term memory (LSTM)~\cite{LSTM} and Q-learning~\cite{Q_learning} to minimize the delay of the mission critical services. 

% Intro - 4 : In this paper, ~ 그래서. 이 논문은 이러한 multi-tier HetVNet에서 CARA 최적화 문제를 MADDPG를 활용해 VUE들의 QoS (downlink throughput) 를 만족하면서 wireless commun.을 할 수 있는 resource management policy 학습 solution을 제시한다.
However, the traditional DRL based learning approaches mostly assumed single-agent systems, which are hard to be applied in practice. A user may fail to learn an optimal policy because of the partially observable and non-stationary environment, which is caused by actions of neighboring users. In this paper, we solve the joint CARA problem of HetVNet with a QoS guaranteeing \textit{MADDPG}~\cite{MADDPG} based approach. Based on the MADDPG strategy, multiple VUEs can learn their own policy to solve the CARA problem in a cooperative manner. Simulation results show that the proposed method outperforms other DRL methods in terms of DL throughput.

% Intro - 5 : The rest of this paper is organized as follows.~
The rest of this paper is organized as follows. The system model of HetVNet and CARA problem are presented in Sec.~\ref{sec:sec2}. Based on the network architecture and problem definition, multi-agent DRL based solution for the joint CARA problem is proposed in Sec.~\ref{sec:sec3}. Sec.~\ref{sec:sec4} shows the results of performance evaluation. Lastly, Sec.~\ref{sec:sec5} concludes this paper.

\section{System Model and problem definition}\label{sec:sec2}
\begin{figure}[t]
    \centering
        \includegraphics[width =0.9\linewidth, height = 5cm ]{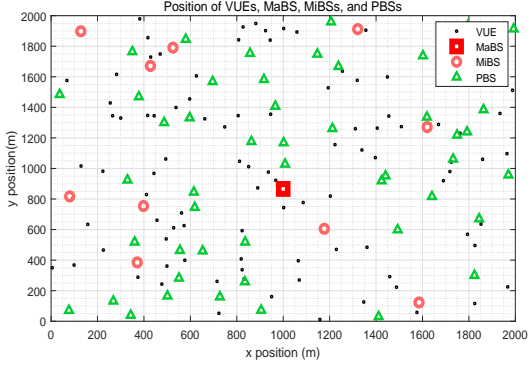}
        %\vspace{-3mm}
    \caption{Network layout with a MaBS, MiBSs, PBSs, and VUEs.}
    \label{fig:NetLayout}
    \vspace{-3mm}
\end{figure}

This section presents the HetVNet system model as Fig.~\ref{fig:NetLayout} and define the joint CARA problem. The 3-tier HetVNet system consists of a macro-cell base station (MaBS), micro-cell base stations (MiBSs), and mmWave-enabled pico-cell base stations (PBSs). In addition, the CARA problem is solved in a way that each VUE cooperatively associates with base station and the wireless resource is efficiently allocated to each VUE so that the downlink throughput of VUEs in the HetVNet can be satisfied.
\subsection{System Model}
The 3-tier HetVNet consists of $K_{a}$ MaBSs, $K_{i}$ MiBSs, and $K_{p}$ PBSs among $K$ base stations, where $K = K_{a} + K_{i} + K_{p}$. In addition, there are $N$ VUEs in the HetVNet. The set of base station is denoted as $\mathcal{B}$, where $\mathcal{B} = \big\{b_{1}, \cdots, b_{K_{a}}, \cdots, b_{K_{a}+K_{i}}, \cdots, b_{K_{a}+K_{i}+K_{p}}\big\}$. To simply note the PBS and other base stations, the set of MaBS and MiBSs is $\mathcal{B}_{\alpha}$, where $\mathcal{B}_{\alpha} = \big\{b_{1}, \cdots, b_{K_{a}}, \cdots, b_{K_{a}+K_{i}}\big\}$. The set of PBSs is denoted as $\mathcal{B}_{\beta}$, where $\mathcal{B}_{\beta} = \big\{b_{K_{a}+K_{i}+1}, \cdots, b_{K_{a}+K_{i}+K_{p}}\big\}$. In addition, each VUE can associate with only one base station during a time slot and it is equipped with one antenna. Suppose that a binary vector $l_{i}^{k}$ represents the cell association information of $i$-th VUE with a base station among $\mathcal{B}$. Then, the vector $l_{i}^{k}(t)$ can be denoted as $l_{i}^{k}(t) = (l_{i}^{1}(t), \cdots, l_{i}^{K}(t))$, where $k \in [1, K]$ and $i \in [1, N]$. If the $i$-th VUE associates with $k$-th base station, then the $l_{i}^{k}(t)$ is set to 1. Otherwise, the value is set to 0. Then, the vector $l_{i}^{k}(t)$ can be represented as:
\begin{equation}
    \label{eq:eq1}
    \sum_{k=1}^{K} l_{i}^{k}(t) \leq 1, \forall i \in [1, N].
\end{equation}
In addition, each VUE can utilize \textit{carrier aggregation}, which combines multiple subchannels of associated base station. However, for fair resource access, each VUE is limited to utilize the spectrum at most $\bar{c}$. We assume that MaBSs and MiBSs share $S$ orthogonal channels and PBSs operate on $P$ mmWave channels. The $C$ represents the set of orthogonal channels and it can be denoted as $C = \big\{c_{1}, \cdots, c_{S}\big\}$. In addition, the set of mmWave channel $M$ can be signified as $M = \big\{m_{1}, \cdots, m_{\mathcal{P}}\big\}$. Then, resource allocation vector between $i$-th VUE and $j$-th MaBS or MiBS can be denoted as $f_{i}^{j}(t)$, where $f_{i}^{j}(t) = (c_{1}^{j}(t), \cdots, c_{S}^{j}(t))$, $i \in [1, N]$ and $j \in [1, K_{a}+K_{i}]$. If the VUE use the $u$-th channel among the $f_{i}^{j}(t)$, then the $f_{i}^{u}(t)$ is set to 1. Otherwise, it is set to 0. However, if the VUE associates with $o$-th PBS, the resource allocation vector between $i$-th VUE and $o$-th PBS can be denoted as $f_{i}^{o}(t) =  (m_{1}^{o}(t), \cdots, m_{\mathcal{P}}^{o}(t))$, where $i \in [1, N]$ and $o \in [K_{a}+K_{i}+1, K_{a}+K_{i}+K_{p}]$. The criteria of setting the value of $f_{i}^{o}(t)$ is as same as the way to $f_{i}^{j}(t)$. Then, the resource allocation toward $i$-th VUE can be denoted as:
\begin{equation}
    \label{eq:eq2}
        \sum_{o=K_{a}+K_{i}+1}^{K} f_{i}^{o}(t) + \sum_{j=1}^{K_{a}+K_{i}} f_{i}^{j}(t) \leq \bar{c}, \forall i \in [1, N].
\end{equation}
Because the MiBSs coexist in the coverage of MaBS, the co-channel interference should be taken into account. In practice, the transmit power value can be defined with a finite number. Suppose that $p_{i, j}^{c}(t)$ stands for the possible transmit power level vector of MaBS or MiBS on the shared spectrum, $p_{i, j}^{c}(t) = (p_{i, j}^{c_{1}}(t), \cdots, p_{i, j}^{c_{S}}(t))$ represent the transmit power per each channel in $C$. In addition, each VUE is assumed to measure instantaneous channel gain $h_{i}^{j}(t)$, where $i \in [1, N]$ and $j \in [1, K]$. Then, the signal-to-interference-plus-noise ratio (SINR) at $i$-th VUE, which is associated with $k$-th base station among $\mathcal{B}$ (i.e., $b_{k}$) using channel $f_{i}^{j}(t)$ or $f_{i}^{o}(t)$, is as follows (denoted by $\Psi_{i, k}^{C, M}(t)$):
\begin{equation}
    \label{eq:eq3}
    \resizebox{1.0\hsize}{!}{$\frac{h_{i}^{k}(t)f_{i}^{j}(t)p_{i, j}^{c}(t) + h_{i}^{k}(t)f_{i}^{o}(t)p_{i,j}^{m}(t)}{\sum_{v}^{\mathcal{B}_{\alpha} - \big\{b_{k}\big\}} h_{i}^{v}(t)f_{i}^{j}(t)p_{j, i}^{c}(t) + \sum_{v}^{\mathcal{B}_{\beta}- \big\{b_{v}\big\}}h_{i}^{k}(t)f_{i}^{o}(t)p_{j, i}^{m}(t) + WN_{0}},
    $}
\end{equation}
where $W$ is the bandwidth of a channel, $N_{0}$ stands for the noise power, $v \in \mathcal{B}, i \in [1, N], j \in [1, K_{a} + K_{i}]$, and $o \in [K_{a}+K_{i}+1, K_{a}+K_{i}+K_{p}]$. Based on Eq.~(\ref{eq:eq3}), the DL throughput of $i$-th VUE, which is denoted as $\zeta_{i}(t)$ can be:
\begin{equation}
    \label{eq:eq4}
    \zeta_{i}(t) = \sum_{k=1}^{K} l_{i}^{k}(t) \sum_{\forall z \in C \cup M} W\log_{2}(1 + \Psi_{i, k}^{z}(t)).
\end{equation}
\subsection{CARA Problem Formulation}
Based on the aforementioned system model, the joint CARA problem can be defined in a way that the VUEs are satisfied with minimum QoS baseline $\chi$, while they cooperatively associate with base stations and utilize wireless resource, i.e.,
\begin{equation}
    \label{eq:eq5}
    \sum_{k=1}^{K}l_{i}^{k}(t)\sum_{\forall z \in C \cup M}\Psi_{i, k}^{z}(t) \geq \chi.
\end{equation}
Considering the transmit power of $k$-th bsae station toward $i$-th VUE, the power-aware cost can be changed as:
\begin{equation}
    \label{eq:eq6}
    \kappa_{i}(t) = \sum_{k=1}^{K}\rho l_{i}^{k}(t)\left[\sum_{o=1}^{K_{p}}f_{i}^{o}(t)p_{i, j}^{c}(t) + \sum_{j=1}^{K_{a}+K_{i}}f_{i}^{j}(t)p_{i, j}^{m}(t)\right],
\end{equation}
where the $\rho$ stands for the cost of unit power level. Overall, the total revenue $\Lambda_{i}(t)$ of the $i$-th VUE in the HetVNet system can be formulated as:
\begin{equation}
    \label{eq:eq7}
    \Lambda_{i}(t) = \eta \zeta_{i}(t) - \kappa_{i}(t),
\end{equation}
where the $\eta$ stands for the positive profit of each channel capacity. Hence, the objective of joint CARA problem is to optimize the expected total return of Eq.~(\ref{eq:eq7}) under Eq.~(\ref{eq:eq5}). The expected total return of the revenue of the $i$-th VUE can be denoted as $\Omega_{i}(t)$, i.e.,
\begin{equation}
    \label{eq:eq8}
    \Omega_{i}(t) = \sum_{t=1}^{+\infty} \gamma^{t-1}\Lambda_{i}(t),
\end{equation}
where the $\gamma \in [0, 1)$ is the discounting factor in reinforcement learning to represent the uncertainty of future revenue. Throughout the Eq.~(\ref{eq:eq1}) to Eq.~(\ref{eq:eq8}), VUEs and base stations dynamically transit their resource utilization state and action, which is highly combinatorial and intractable to optimize. In this regard, $policy$ optimization based multi-agent DRL is derived to solve the joint CARA problem of HetVNet. 

\section{Multi-agent DRL for cooperative CARA problem in HetVNet}\label{sec:sec3}
Throughout multiple interactions with the HetVNet (i.e., the environment), each VUE accumulates its own experiences, which is paired with $(o_{i}(t), a_{i}(t), r_{i}(t), o_{i}(t+1))$. The $o_{i}(t) \in \mathcal{O}$ stands for the local observation of $i$-th VUE at time slot $t$. The $a_{i}(t) \in \mathcal{A}$ denotes the action of the VUE. Lastly, the $r_{i}(t)$ signifies the temporal difference reward of the VUE. The aforementioned traditional single-agent approaches to solve the joint CARA problem are not capable of learning the cooperative spectrum access policy of VUEs, because of the non-stationary environment. The $r_{i}(t)$ may differ from the same $o_{i}(t)$ and $a_{i}(t)$ in the set of experience pairs, because the observation $o_{i}(t)$ of $i$-th VUE only contains local information. That is, the VUE only has local information of the HetVNet so that states and actions of other VUEs, which impact on the VUE's \textit{reward}, may differ even the same local observation and action of VUE. Thus, to solve the joint CARA problem with multiple VUEs, policy updating procedure of a VUE should take into account actions of other VUEs, rather updating the policy only with its own action. Therefore, the multi-agent approach is more suitable for optimizing the policies of VUEs to solve the joint CARA problem in HetVNet.

\subsection{Preliminaries of Reinforcement Learning}
The reinforcement learning agent learns how to act (i.e., policy) in a sequential decision making problem through interactions between its environment. The decision making problem can be modeled as a Markov decision process (MDP), which is the pairs of $(o_{i}(t), a_{i}(t), r_{i}(t), o_{i}(t+1))$. The observation space $\mathcal{O}$ stands for the set of possible observations of VUEs and the action space $\mathcal{A}$ denotes the set of possible actions of them. The agent aims to optimize its policy $\mu_{\theta_{i}}$, which is parameterized with $\theta_{i}$, i.e., $a_{i}(t) = \mu_{\theta_{i}}(o_{i}(t))$. The policy updating procedure changes the parameter $\theta$ in a way the expected total return of the agent with respect to $a_{i}(t)$ for given $o_{i}(t)$ is improved. The value of action for sequential observations (i.e., state $s$) is measured with action-value function, or Q-function, to evaluate the expected total return per action. The Q-function can be denoted as follows:
\begin{eqnarray}
    Q^{\mu}(s, a) &=& \mathbb{E}[\Omega_{i}(t)|s=s_{i}(t), a=a_{i}(t)]\\
    &=& \mathbb{E}_{s^{'}}[r(s, a)+\gamma\mathbb{E}_{a^{'}\thicksim \mu}[Q^{\mu}(s^{'}, a^{'})]].
\end{eqnarray}

\subsection{MADDPG Approach on CARA}
Here, the multi-agent deep reinforcement learning strategy is presented to solve the joint CARA problem. In DRL, deep neural network (DNN) is utilized to build the learning agent. The DNN takes a role of a non-linear approximator to obtain the optimal policies $\mu^{*}$ VUEs. Suppose that $\mu = \big\{\mu_{1}, \cdots, \mu_{N}\big\}$ be the set of all agent policies and $\theta = \big\{\theta_{1}, \cdots, \theta_{N}\big\}$ is the parameter set of corresponding policy. Based on estimation of Q-function for each possible action, VUEs update their own policy. The MADDPG is policy gradient based off-policy \textit{actor-critic} algorithm~\cite{MADDPG}, where the objective function $\mathcal{J}(\theta)$ is expected reward, i.e., $\mathcal{J}(\theta_{i}) = \mathbb{E}[\Omega_{i}(t)]$. That is, the optimal policy of $i$-th VUE can be represented as $\mu_{\theta_{i}}^{*} = \arg \max_{\mu_{\theta_{i}}}\mathcal{J}(\theta_{i})$. To optimize the objective function, the gradient of the objective function is calculated with respect to $\theta_{i}$ as:
\begin{equation}
    \label{eq:eq9}
    \nabla_{\theta_{i}}\mathcal{J}(\mu_{i}) = \mathbb{E}_{\textbf{x}, a \thicksim \mathcal{D}}[\nabla_{\theta_{i}}\mu_{i}(a_{i}|o_{i})\nabla_{a_{i}}Q_{i}^{\mu}(\textbf{x}, a_{1}, \cdots, a_{N})],
\end{equation}
where the $\textbf{x} = (o_{1}, \cdots, o_{N})$, $Q_{i}^{\mu}(\textbf{x}, a_{1}, \cdots, a_{N})$ is a centralized action-value funciton, and replay buffer $\mathcal{D}$. The $\mathcal{D}$ contains transition tuples $(\textbf{x}, a, r,  \textbf{x}^{'})$, where $a = (a_{1}, \cdots, a_{N})$ and $r = (r_{1}, \cdots, r_{N})$. The centralized action-value function $Q_{i}^{\mu}$ is updated for minimizing the loss function \eqref{eq:eq10}:
\begin{equation}
\label{eq:eq10}
    \mathcal{L}(\theta_{i}) = \mathbb{E}_{\textbf{x}, a, r, \textbf{x}^{'}}[(Q_{i}^{\mu}(\textbf{x}, a_{1}, \cdots, a_{N})-y)^2],
\end{equation}
where $y = r_{i} + \gamma Q_{i}^{\mu^{'}}(\textbf{x}^{'}, a_{1}^{'}, \cdots, a_{N}^{'})|_{a_{j}^{'}=\mu_{j}^{'}(o_{j})}$. The $\mu^{'} = \big\{\mu_{\theta_{1}^{'}}, \cdots, \mu_{\theta_{N}^{'}}\big\}$ stands for the target policies with delayed parameters $\theta_{i}^{'}$. In addition, the MADDPG is \textit{actor-critic} based algorithm, where the \textit{actor} takes a role of making sequential decisions over time slots, while the \textit{critic} evaluates the behavior of the \textit{actor}. Each VUE agent consists of the \textit{actor} and \textit{critic} with behavior network and target network. The \textit{actor} updates the behavior policy network and periodically update the target policy network by utilizing gradient ascent updating manner on the $\mathcal{J}(\theta)$ with Eq.~(\ref{eq:eq9}). Similarly, the \textit{critic} updates the behavior Q-function and periodically updates the target Q-function in a way that minimizes the loss function in Eq.~(\ref{eq:eq10}). The VUEs have such \textit{actor} and \textit{critic} to optimize their own policy to behave cooperatively, while they update their \textit{critic}'s Q-function to reasonably evaluate the actions. To be more specific, the optimization objective of such policy gradient approach is updating the $\theta$ of target network, which makes the VUEs actually how to act. The value of neural network of target network of \textit{actor} is fixed for a number of iterations, while the weights of neural network of behavior network of \textit{actor} are updated.

That is, the multi-agents in HetVNet observe their local information and aim to act in a way that maximize their expected total return. They can stably update the policy parameter $\theta$ even though the local information and interactions between other VUEs and HetVNet. In other words, the environment is stationary even as the policies change. Suppose that $P$ stands for the state transition probability, $P(s'|s, a_{1}, \cdots, a_{N}, \mu_{1}, \cdots, \mu_{N} = P(s'|s, a_{1}, \cdots, a_{N}) = P(s'|s, a_{1}, \cdots, a_{N}, \mu_{1}^{'}, \cdots, \mu_{N}^{'})$ for any $\mu_{i} \neq \mu_{i}^{'}$. Therefore, because the state transition probability from $s$ to $s^{'}$ of VUE is same even though the behavior policy and target policy are mutually different.
\subsubsection{State Space}
The state space of each VUE in HetVNet is defined with two-folds: QoS satisfaction and accumulative DL throughput variation. The state of $i$-th VUE in terms of QoS $s_{i}^{qos}(t)$ is set to 1 if $\Psi_{i, k}^{C, M}(t) \geq \chi$, or is set to 0 otherwise. In addition, the DL throughput of current time slot is compared to previous one to decide the $s_{i}^{dl}(t)$. The $s_{i}^{dl}(t)$ is set to 1 if the DL throughput of current time slot is higher than previous one, or is set to 0 otherwise. Therefore, the state space of VUEs $s(t)$ can be defined as $s(t) = (s_{1}(t), \cdots, s_{N}(t)) = \big\{(s_{1}^{qos}(t), s_{1}^{dl}(t)), \cdots, (s_{N}^{qos}(t), s_{N}^{dl}(t))\big\}$.
\subsubsection{Action Space}
The VUE decides actions to choose for every time slot. It firstly decides what kind of base station to associate between MaBS/MiBS and PBS. In addition, it chooses which channels to utilize for communication. Thus, the action space of VUEs can be defined as $a(t) = (a_{1}(t), \cdots, a_{N}(t)) = \big\{(l_{1}^{k}(t), f_{1}^{o}(t), f_{1}^{j}(t)), \cdots, (l_{N}^{k}(t), f_{N}^{o}(t), f_{N}^{j}(t))\big\}$, where $\forall i \in [1, N], o \in [K_{a}+K_{i}+1, K], j \in [1, K_{a}+K_{i}]$. As the number of PBSs is increased, the action space exponentially grow so that it is intractable to solve the joint CARA problem with traditional approaches.
\subsubsection{Reward Structure}
The immediate reward of $i$-th VUE can be denoted as $\mathcal{R}_{i}(t)$ and it can be computed based on the interaction between VUEs and the HetVNet, i.e., $((s_{1}(t), a_{1}(t)), \cdots, (s_{N}(t), a_{N}(t)))$. Then, the $\mathcal{R}_{i}(t)$ can be:
\begin{equation}
    \label{eq:eq11}
    \mathcal{R}_{i}(t) = \Lambda_{i}(t) - \Upsilon_{i}.
\end{equation}
Note that the $\Upsilon_{i}$ stands for the failure penalty of $i$-th VUE, which is took into account for the calculation of the reward when the VUE fails to associate with a base station or it cannot access any wireless spectrum.
\subsection{Algorithm Description}
The MADDPG based algorithm to solve the joint CARA problem is presented in this section. The detailed description of the algorithm is as follows:
\begin{algorithm}[t]
\small
\label{algorithm1}
    Initialize the weights of \textit{actor} and \textit{critic} networks~\\
    Initialize a random process $\mathcal{N}$ for exploration of action~\\
    Receive the initial state \textbf{x}~\\
    \For{t = 1 to $\mathcal{E}$}{
        $\triangleright$ Each VUE selects a cell association and resource utilization action $a_{i} = \mu_{\theta_{i}}(o_{i}) + \mathcal{N}_{t}$ based on the exploration and policy~\\ 
        $\triangleright$ Execute actions $a = (a_{1}(t), \cdots, a_{N}(t))$~\\
        $\triangleright$ Observe rewards $\mathcal{R}(t)$ and new state $\textbf{x}^{'}$~\\
        $\triangleright$ Store $(\textbf{x}, a, \mathcal{R}(t), \textbf{x}^{'})$ in $\mathcal{D}$~\\
        $\triangleright$ $\textbf{x} \leftarrow \textbf{x}^{'}$~\\
        \For{agent $i$ = 1 to $N$}{
            $\triangleright$ Sample a random minibatch of $\mathcal{V}$ samples $(\textbf{x}^{j}, a^{j}, \mathcal{R}^{j}, \textbf{x}^{'j})$ from $\mathcal{D}$ \\
            $\triangleright$ Set $y^{j} = \mathcal{R}_{i}^{j} + \gamma Q_{i}^{\mu^{'}}(\textbf{x}^{'j}, a_{1}^{'}, \cdots, a_{N}^{'})|_{a_{k}^{'} = \mu_{k}^{'}(o_{k}^{'})}$ \\
            $\triangleright$ Update behavior critic by minimizing the loss $\mathcal{L}(\theta_{i}) = \frac{1}{\mathcal{V}}\sum_{j}(y^{j}-Q_{i}^{\mu}(\textbf{x}^{j}, a_{1}^{j}, \cdots, a_{N}^{j}))^2$ \\
            $\triangleright$ Update behavior actor using the sampled policy gradient:
            $\nabla_{\theta_{i}}\mathcal{J} \approx 
                \frac{1}{\mathcal{V}}\sum_{j}\nabla_{\theta_{i}}\mu_{i}(o_{i}^{j})\nabla_{a_{i}}Q_{i}^{\mu}(\textbf{x}^{j}, a_{1}^{j}, \cdots, a_{i}, \cdots, a_{N}^{j})|_{a_{i}=\mu_{i}(o_{i}^{j})}$
        }
        $\triangleright$ Update the \textit{target} network parameters of each VUE:
        $\theta_{i}^{'} \leftarrow \tau\theta_{i} + (1-\tau)\theta_{i}^{'}$
    }
    \caption{MADDPG algorithm for joint CARA problem}
\end{algorithm}
\begin{itemize}
    \item First, the parameters of the \textit{actor} and \textit{critic} network, which activate and evaluate the action of VUEs, are initialized (line 1--3).
    \item Next, for $\mathcal{E}$ iterations, following procedures are conducted to update the target network parameters of VUEs. Given the initial state $\textbf{x}$, each VUE selects its action based on the exploration noise and its own policy (line 5). After the actions of each VUE are conducted, then the actions are activated by the VUEs (line 6). Next, the HetVNet interacts with the VUEs and returns corresponding rewards and next states (line 7). Then, each VUE observes the state transition pair and stores in the replay buffer $\mathcal{D}$, which contains the experiences of VUEs (line 8). Then, the episodic state $\textbf{x}$ is changed to the next $\textbf{x}^{'}$ (line 9).
    \item Throughout the episode, each VUE conducts following procedures to update their \textit{actor} and \textit{critic} networks. At first, an $i$-th VUE samples a random minibatch of $\mathcal{V}$ samples among $\mathcal{D}$ (line 11). Note that the superscript $j$ stands for the approximation of other VUEs of $i$-th VUE. Then, the target value of Q-function $y^{j}$ is set (line 12). By minimizing the difference between $y^{j}$ and $Q_{i}^{\mu}(\textbf{x}^{j}, a_{1}^{j}, \cdots, a_{N}^{j})$ among $\mathcal{V}$ samples, the $\theta$ of behavior \textit{critic} is updated (line 13). Similarly, the $\theta$ of $\mathcal{J}$ of behavior \textit{actor} is updated with the gradient to optimize the policy parameter $\theta$ (line 14). Note that the policy update is based on gradient ascent calculation.
    \item Lastly, after all VUEs update their behavior networks, the target network parameters are updated under the concept of \textit{soft update} (line 16).
\end{itemize}% End of section 3.

\section{Performance evaluation}\label{sec:sec4}
In this section, we provide the performance evaluation setting of MADDPG algorithm to solve the joint CARA problem. We considered 1 MaBS, 10 MiBSs, 50 PBSs, and 100 VUEs in HetVNet as Fig.~\ref{fig:NetLayout}. In case of cell coverage region of each base station, the radius of MaBS is set to 3000m, while the MiBS and PBS are set to 500m and 100m, respectively. The transmit powers of MaBS, MiBS, and PBS are set to 40\textit{dBm}, 35\textit{dBm}, and 20\textit{dBm}, respectively. The $\mathcal{S}$ is set to 30, while the $\mathcal{P}$ is set to 5. The channel bandwidth of MaBS/MiBS is set to 180\textit{kHz} and the DL center frequency is 2\textit{GHz}. Meanwhile, the channel bandwidth of PBS is set to 800\textit{MHz} and the DL center frequency is 28\textit{GHz}. The path loss of MaBS and MiBS is set to $34 + 40\log(d)$ and the PBS's one is set to $37 + 30\log(d)$. All the failure cost $\Upsilon_{i}$ of $\mathcal{R}_{i}(t)$ is set to $10^{-2}$ and the base line of QoS $\chi$ is set to 7\textit{dBm}. The noise power $N_{0}$ is set to -175\textit{dBm}/\textit{Hz} and the $\rho$ is set to $10^{-3}$. The MADDPG model is composed with two-layerd fully connected neural networks with 64 and 32 neurons, respectively. The hyperparameter of the model can be summarized as Table~\ref{tab:tab1}.
\begin{table}[h]%\\
\caption{Hyperparameter of MADDPG model}
\small
\label{tab:tab1}
\begin{center}
	\centering
	\begin{tabular}{r||r}
    \toprule[1.0pt]\centering
     Parameter & Value\\
    \midrule
    \midrule
    Total episode $\mathcal{E}$ & 500\\
    Time step $T$ & 100\\
    Minibatch size & 64\\
    Discounting factor $\gamma$ & 0.95\\
    Initial epsilon & 0.9\\
    Learning rate $\delta$ & 0.05\\
    Size of $\mathcal{D}$ & 1000\\
    Optimizer & AdamOptimizer\\
    Activation function & ReLU\\
    \bottomrule[1.0pt]
	\end{tabular}
\end{center}
\end{table}
\begin{figure}[t]
    \centering
        \includegraphics[width =0.99\linewidth, height = 6cm ]{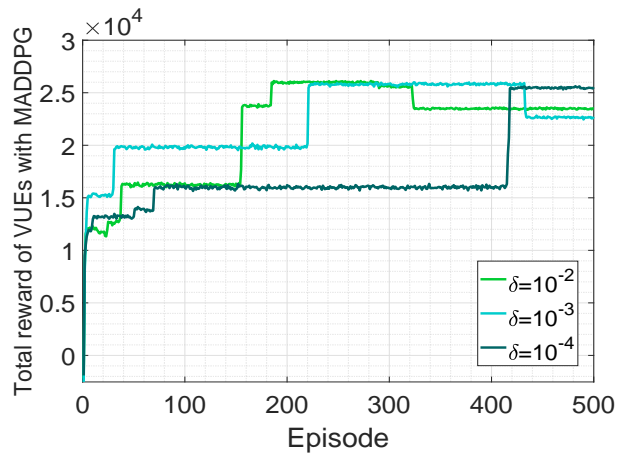}
        %\vspace{-3mm}
    \caption{Convergence rate comparison for different learning rate $\delta$ with MADDPG algorithm.}
    \label{fig:fig2}
    \vspace{-3mm}
\end{figure}

Firstly, the learning curve of the MADDPG to solve the joint CARA problem is as Fig.~\ref{fig:fig2}. The convergence points of each learning model are slightly different to each other. It shows that the required episode to get the converged performance is decreased as the learning rate is smaller. 

\begin{figure}[t]
    \centering
        \includegraphics[width =0.99\linewidth, height = 6cm ]{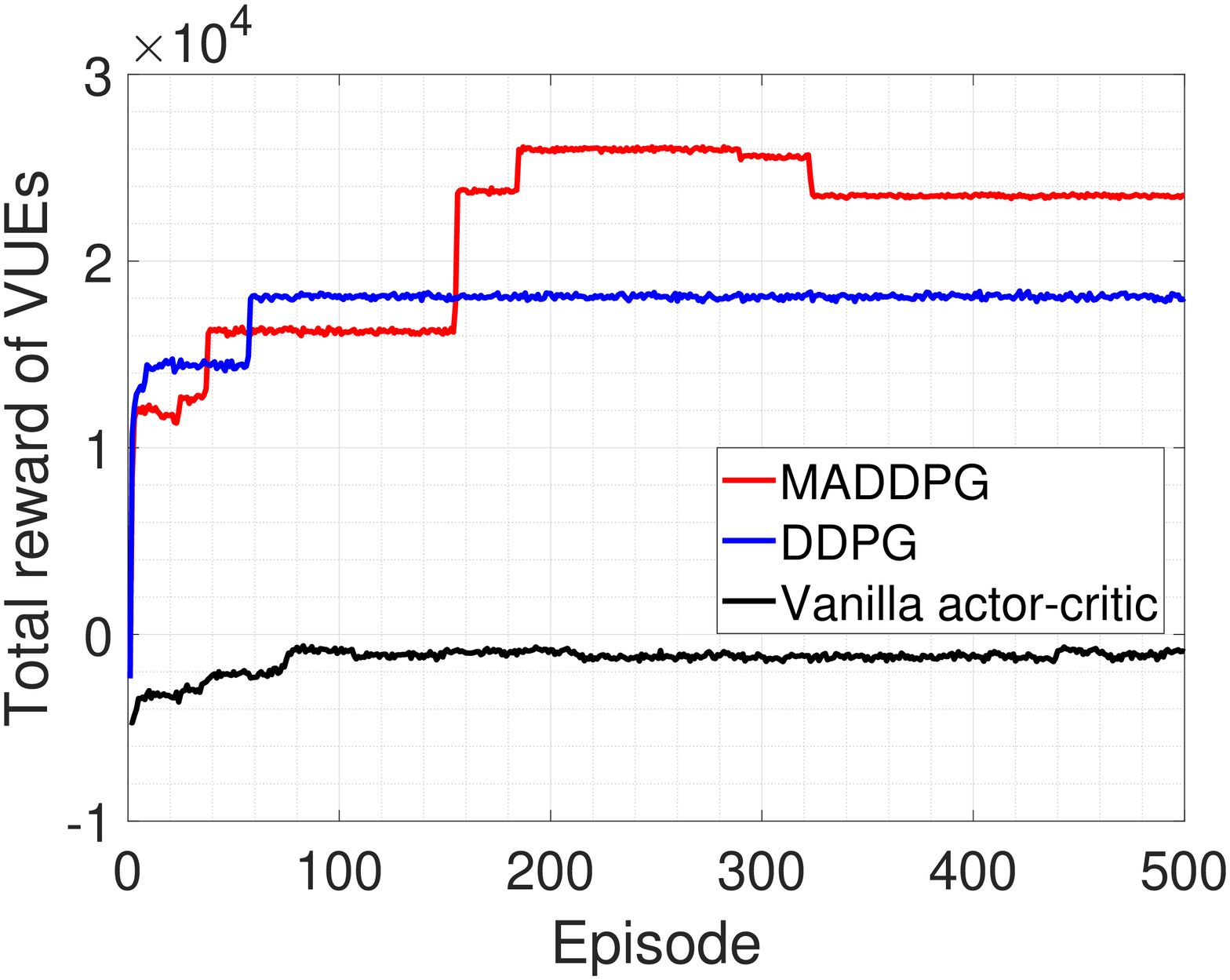}
        %\vspace{-3mm}
    \caption{Performance comparison on joint CARA problem in 3-tier HetVNet.}
    \label{fig:fig3}
    \vspace{-3mm}
\end{figure}

Next, the performance of MADDPG strategy is compared with other policy gradient (PG) algorithms, i.e., vanilla actor--critic and DDPG approaches. Note that the vanilla actor--critic is a baseline algorithm among PG algorithms. As shown in Fig.~\ref{fig:fig3}, the vanilla actor--critic approach almost fails to solve the joint CARA problem, so that each VUEs greedily access the wireless spectrum and suffer from the collision, while DDPG and MADDPG strategies showed much higher performance. However, because of the non-stationary problem of DDPG, the total reward of VUEs trained by the MADDPG is higher than that of DDPG.

\begin{figure}[t]
    \centering
        \includegraphics[width =0.99\linewidth, height = 6cm ]{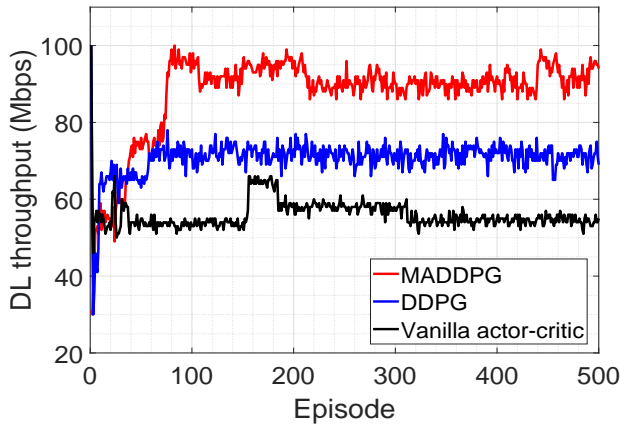}
        %\vspace{-3mm}
    \caption{DL throughput of each algorithm in HetVNet.}
    \label{fig:fig4}
    \vspace{-3mm}
\end{figure}

Finally, the performance of average DL throughput of VUEs among HetVNet is provided as Fig.~\ref{fig:fig4}. Considering Fig.~\ref{fig:fig3} and Fig.~\ref{fig:fig4}, the MADDPG strategy to solve the joint CARA problem learned policies of VUEs in a way that cooperatively associate with base stations and utilize wireless spectrums (high total reward of VUEs as Fig.~\ref{fig:fig3} and high DL throughput as Fig.~\ref{fig:fig4}). Although the DDPG-based solution showed somewhat lower performance than that of MADDPG, it is still showed to learn cooperative policies as Fig.~\ref{fig:fig3}. However, the vanilla actor--critic approach showed the lowest total reward of VUEs as Fig.~\ref{fig:fig3} and DL throughput as Fig.~\ref{fig:fig4}, which stands for the vanilla actor--critic approach learned selfish association and resource utilization policies under non-stationary environment setting. In conclusion, the MADDPG strategy was successful to learn cooperative policies to solve joint CARA problem in considered HetVNet. 

\section{Conclusion}\label{sec:sec5}
In this paper, we proposed multi-agent DRL approach to solve the joint CARA problem in HetVNet. Because of the non-stationary problem and NP-hard property, the traditional approaches including single agent RL methods were limited to solve the problem. However, the proposed MADDPG strategy showed a near optimal solution with a small number of iterations and the achieved better DL throughput performance compared to other reinforcement learning methods.

\section*{Acknowledgment}
This research was supported by the National Research Foundation of Korea (2019R1A2C4070663); and also by Institute for IITP grant funded by MSIT (No.2018-0-00170, Virtual Presence in Moving Objects through 5G).
J. Kim is the corresponding author of this paper.

\end{document}